# Tunable linearity of high-performance vertical dual-gate vdW phototransistor

*Jinpeng Xu, Xiaoguang Luo, Siqi Hu, Xi Zhang, Dong Mei, Fan Liu, Nannan Han, Dan Liu, Xuetao Gan\*, Yingchun Cheng\*, Wei Huang\**

J. Xu, Dr. X. Luo, X. Zhang, D. Mei, F. Liu, Dr. N. Han, Dr. D. Liu, Prof. Y. Cheng, Prof. W. Huang
Frontiers Science Center for Flexible Electronics (FSCFE), Shaanxi Institute of Flexible Electronics (SIFE) & Shaanxi Institute of Biomedical Materials and Engineering (SIBME)
Northwestern Polytechnical University, Xi'an 710129, China
E-mail: iamyccheng@njtech.edu.cn
E-mail: iamwhuang@nwpu.edu.cn

S. Hu, Prof. X. Gan
MOE Key Laboratory of Material Physics and Chemistry under Extraordinary Conditions, and Shaanxi Key Laboratory of Optical Information Technology
School of Physical Science and Technology
Northwestern Polytechnical University, Xi'an 710129, China
E-mail: xuetaogan@nwpu.edu.cn

Prof. Y. Cheng, Prof. W. Huang
Key Laboratory of Flexible Electronics & Institute of Advanced Materials, Jiangsu National Synergetic Innovation Center for Advanced Materials
Nanjing Tech University, Nanjing 211816, China

**Abstract**

Layered two-dimensional (2D) semiconductors have been widely exploited in photodetectors due to their excellent electronic and optoelectronic properties. To improve their performance, photogating, photoconductive, photovoltaic, photothermoelectric, and other effects have been used in phototransistors and photodiodes made with 2D semiconductors or hybrid structures. However, it is difficult to achieve the desired high responsivity and linear photoresponse simultaneously in a monopolar conduction channel or a *p-n* junction. Here we present dual-channel conduction with ambipolar multilayer $WSe_2$ by employing the device concept of dual-gate phototransistor, where *p*-type and *n*-type channels are produced in the same semiconductor using opposite dual-gating. It is possible to tune the photoconductive gain using a vertical electric field, so that the gain is constant with respect to the light intensity – a linear photoresponse, with a high responsivity of ~$2.5\times10^4$ A $W^{-1}$. Additionally, the 1/f noise of the device is kept at a low level under the opposite dual-gating due to the reduction of current and carrier fluctuation, resulting in a high detectivity of ~$2\times10^{13}$ Jones in the linear photoresponse regime. The linear photoresponse and high performance of our dual-gate $WSe_2$ phototransistor offer the possibility of achieving high-resolution and quantitative light detection with layered 2D semiconductors.

## 1. Introduction

Layered two-dimensional (2D) semiconductors are considered as candidates for next-generation electronic and optoelectronic technologies, due to qualities such as their high carrier mobility, bandgap tunability (by the number of layers or strain), and large absorption coefficient.[1, 2] The nature of atomic thin thickness and hanging-bonds-free surfaces makes 2D semiconductors available for the device applications that constructed and integrated with the distinct form of van der Waals contacts.[3] Photodetectors such as phototransistors and photodiodes, which convert light to electrical signals, are widely made from 2D semiconductors.

Both unipolar and ambipolar 2D semiconductors are frequently considered as the single active material for phototransistors.[4-6] High responsivity with high external quantum efficiency (EQE) can be achieved readily by the photogating effect if there are abundant trap states inside the channel materials[7-9] or at the interfaces.[10] Trap states produce a great photoconductive gain which reduces gradually with the increase of light intensity, leading to a sublinear photoresponse (sublinear relationship between photocurrent and light intensity). Taking the typical $n$-type monolayer $MoS_2$ phototransistor as an example, a high responsivity of ~880 A W$^{-1}$ (with EQE ~2×10$^5$ %) is achieved.[11] However, the long lifetime of traps prolongs the response time to the order of 10 s, and so significantly limits the detection speed. Narrow bandgap 2D semiconductors are often chosen for the detection of wide spectral bandwidth, although they exhibit a large parasitic dark current.[12, 13] The detected wavelength can be extended beyond the bandgap through thermal effects (photothermoelectric or bolometric effects) of 2D semiconductors, further sacrificing speed.[14, 15]

In contrast to phototransistors, 2D photodiodes with a *p-n* junction or Schottky junction usually work faster because of the strong built-in electric field.[16-19] It can be predicted that the photoresponse of a typical 2D photodiode is linear due to the absence of photoconductive gain, and the dark current is very weak under reverse bias or even no bias (photovoltaic effect). However, the responsivity is very small and the EQE cannot exceed 100%. Generally, high-resolution and quantitative light detection requires a linear photoresponse with high responsivity, which can be realized by phototransistors if the photoconductive gain is constant with respect to light intensity. Several approaches have been reported recently about the gain tunability of phototransistors, such as adjusting the Fermi level by gate control,[20] using photoactive perovskite dielectrics based on photoinduced ionic migration,[21] and rearranging the carriers with a vertical electric field produced by dual-gating.[22, 23] Despite exhibiting a linear photoresponse with high responsivity, the first two of these approaches are limited by unexceptional detectivity (due to large 1/f noise) and unsatisfactory device stability, respectively. The third approach is more promising, namely the employment of a vertical electric field in an ambipolar semiconductor. With a device concept that combines the operation of phototransistors and photodiodes,[23] the 1/f noise can be reduced under the vertical *p-n* homojunction, and the speed can be improved simultaneously because of enhanced photoconduction in the more conducting channel. As far as we know, this device concept has not yet been employed with layered 2D semiconductors.

In this work, dual-gate phototransistors based on multilayer ambipolar $WSe_2$ (bandgap

~1.4 eV) were fabricated with the dry transfer technique.[24] They combined the functions of phototransistors and photodiodes. The photoconductive gain was tuned with the vertical electric field that was produced by the dual-gating, and it could be controlled using opposite dual-gating to make it constant, thus displaying a linear photoresponse. The experimental results showed a high responsivity of ~2.5×10$^4$ A W$^{-1}$ in the linear photoresponse regime, attributed to the large photoconductive gain of ~10$^5$ from the photogating effect. Carriers were separated by the built-in *p-n* homojunction, reducing the 1/f noise and increasing photoconduction. As a result, a high detectivity of ~2×10$^{13}$ Jones was achieved with a good detection speed (rise time 60 ms and decay time 100 ms). Our results provide a promising route forward for high-resolution and quantitative light detection with layered 2D semiconductors.

## 2. Results and discussion

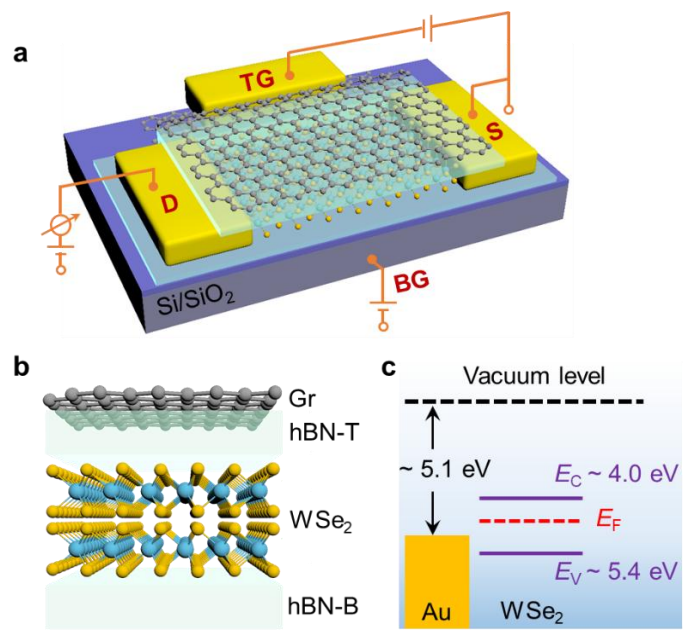

**Figure 1.** Schematic of the dual-gate WSe$_2$ photodetector **a**, and the visualization of the structure **b**, where the WSe$_2$ channel is sandwiched by two hBN nanosheets on the Si/SiO$_2$ substrate. S: source, D: drain, TG: top gate, BG: bottom gate, Gr: graphene, hBN-T/B: top/bottom hBN nanosheet. **c** Energy band diagrams of Au electrode and multilayer WSe$_2$.

To realize a vertical dual-channel configuration in one semiconductor layer, multilayer WSe$_2$ with sufficient thickness (> 20nm) was chosen for the channel material of the transistor. **Figure 1**a and **1**b show the schematic of the dual-gate WSe$_2$ phototransistor on a Si/SiO$_2$ substrate, fabricated with the dry transfer method.[24, 25] Fabrication details can be found in the Experimental Section and **Figure S1**. Exfoliated multilayer WSe$_2$ were sandwiched between hexagonal boron nitride (hBN) nanosheets. The hBN nanosheets behaved as dielectric layers and additionally provided a clean substrate to avoid charge traps and surface bonds at the interface.[26] A few-layer exfoliated graphene, cover the whole WSe$_2$ channel from above, was used as the transparent top gate.

The band structures of the multilayer WSe$_2$ and Au electrode are shown in **Figure 1**c. It is known that the energy levels of the valence band top and the conduction band bottom are ~5.4 eV and ~4.0 eV, respectively.[27] Thus, the Schottky barrier heights for electrons and holes at the Au contacts (work function ~5.1 eV) were predicted theoretically as $\Phi_{ns}$ ~1.1 eV and $\Phi_{ps}$ ~0.3 eV, respectively. The Fermi pinning effect was significantly supressed at the Au/WSe$_2$ van der Waals contact,[28] and the Schottky barriers beneath the contact had a large impact on the conduction polarity of the WSe$_2$ channel. Therefore, predominantly hole transport (i.e., *p*-type behavior) was expected in our phototransistor.

To identify the thickness of each nanosheet, atomic force microscopy (AFM) mapping was carried out around the side of the transistor (white box region in **Figure S1**e), as shown in **Figure S2**a. The thicknesses of WSe$_2$, of the bottom and top hBN nanosheets, and of the graphene, were related to the electrical and optical background of the phototransistor. Raman and photoluminescence (PL) spectra with 532 nm laser excitation were examined to ascertain the quality of the multilayer WSe$_2$ channel, as shown in **Figure S2**b. The two intrinsic Raman peaks at around 249 cm$^{-1}$ and 256 cm$^{-1}$ were assigned to the in-plane vibrational E$^1_{2g}$ mode and out-of-plane vibrational A$_{1g}$ mode, respectively. Another peak at around 308 cm$^{-1}$, assigned to the B$^1_{2g}$ mode, was attributed to the interlayer interaction in multilayer WSe$_2$.[18, 29-31] The PL spectrum had two peaks at around 1.45 eV and 1.6 eV, which arose from the indirect and direct emissions inside multilayer WSe$_2$, respectively.[23, 32-34] Both the Raman and PL spectra implied very high quality interfaces and very few impurities added into the multilayer WSe$_2$ during device fabrication.

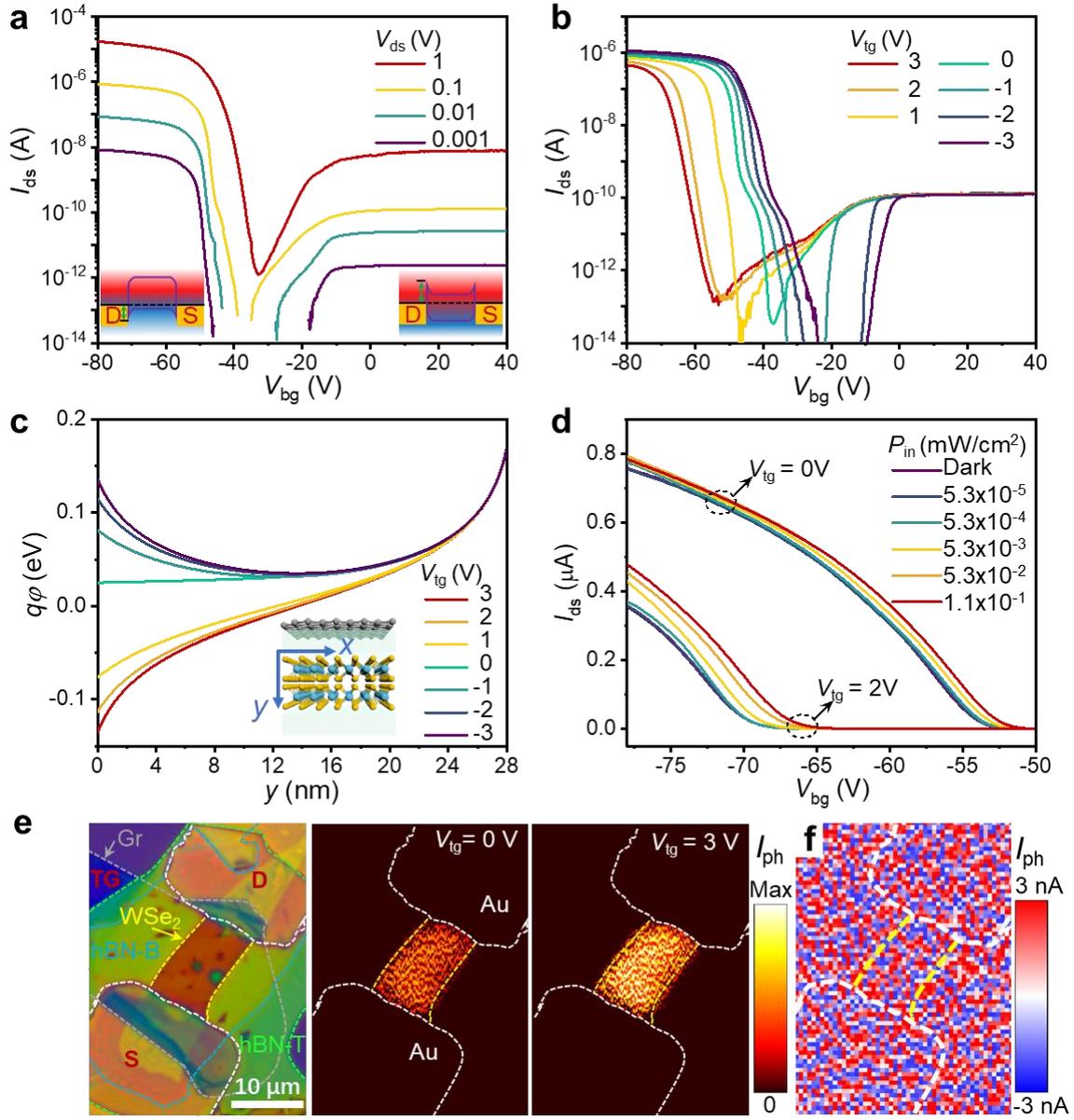

**Figure 2.** Electrical characterization and photoresponse of the dual-gate WSe$_2$ phototransistor. **a** Transfer curves with bottom gating at different drain voltage $V_{ds}$. The insets at left and right corners depict the band structure of the device in the *p*- and *n*-type branches, respectively, with the green arrow lines marking the Schottky barrier height. **b** Transfer curves with dual-gating when top gate voltage $V_{tg}$ increases from −3 V to 3 V with the step of 1 V, the drain voltage $V_{ds}$ = 0.1 V is fixed. **c** The calculated potential profile along the vertical direction of the WSe$_2$ channel with $V_{bg}$ =− 75 V at different $V_{tg}$. The units of eV is regardless of the sign. The inset points out the coordinates defined in the device and the impact from drain voltage is not considered here. **d** Gating response of the phototransistor in dark and illuminated states at different top gate voltages ($V_{tg}$ = 0 and 2 V), where $V_{ds}$ = 0.1 V. **e** Left panel is optical micrograph of a device, two other panels are scanning photocurrent

images at $V_{tg} = 0$ and 3 V when $V_{ds} = 0.1$ V, **f** The scanning photocurrent image at $V_{tg} = 3$ V and $V_{ds} = 0$ V. The bottom gate voltage for photocurrent scanning is $V_{bg} = -70$ V and the laser power intensity is about 30 nW.

The device was then electrically characterized. All the measurements were carried out in a vacuum chamber at room temperature, unless otherwise specified, more details can be found in the Experimental Section. **Figure 2**a shows the transfer curves ($I_{ds}$ versus $V_{bg}$ at a variety of drain voltages $V_{ds}$) of the transistor with the bottom gating only. The threshold voltages of both *p*- and *n*-branches were negative ($<-10$ V), indicating the *n*-doped WSe$_2$ channel that perhaps contributed to the Se vacancies.[35] With the encapsulation of hBN nanosheets, the WSe$_2$ channel was isolated from the ambient effect,[36] and even retained its *n*-type in the air. Typical ambipolar behavior appeared with a prominent hole branch, and the on-off ratios for electrons and holes were ~10$^4$ and ~10$^7$, respectively. This ambipolar transport behavior can be explained by the band diagram of the device. The Fermi level of Au is aligned with the bandgap of WSe$_2$, allowing the injection of electrons and holes into the conduction and valence band, respectively. The nonlinear dependence in the *n*-branch and sublinear dependence in the *p*-branch for low drain voltage (**Figure S3**a) confirmed that the Schottky barrier height for electrons $\Phi_{ns}$ was much larger than that for holes $\Phi_{ps}$,[37] as illustrated in the insets of Figure 2a. Focusing on the *p*-branch, the hole mobility in the linear region of the transfer curves was calculated as ~76 cm$^2$ V$^{-1}$ s$^{-1}$) at $V_{ds} = 0.1$ V, from the equation $\mu = (\Delta I_{ds}/\Delta V_{bg}) \cdot (L/WC_B V_{ds})$, where $\Delta I_{ds}/\Delta V_{bg}$ is the slope of the transfer curve in the linear region, $L$ and $W$ are the channel length and width, respectively, $C_B = 1/(1/C_{SiO_2} + 1/C_{hBN-B})$ is the capacitance of bottom dielectrics, and $C_{hBN-B}$ is the capacitance of bottom hBN nanosheet. The capacitances of SiO$_2$ and bottom hBN were extracted from the relative dielectric constant $\varepsilon$ and thickness $d$, i.e., $C = \varepsilon\varepsilon_0/d$. It was calculated that $C_{\text{SiO}_2} = 12.1$ nF cm$^{-2}$ and $C_{\text{hBN-B}} = 70$ nF cm$^{-2}$ from dielectric constants 3.9 and 3 for SiO$_2$ and hBN, respectively.[32, 26, 38] Notice that the calculated mobility, consistent with the reported value of multilayer WSe$_2$,[38, 39] is actually underestimated due to the Schottky contacts, and could be further improved through Ohmic contacts by choosing suitable electrode materials.[37, 3]

To form the vertical dual-channel, a top gate was applied to our device. **Figure 2**b shows the transfer characteristics at $V_{\text{ds}} = 0.1$ V under the dual-gating. The top gate voltage $V_{\text{tg}}$ was increased from −3 to 3 V in 1 V steps, and the threshold voltages in both *p*- and *n*-branches shifted negatively due to the depletion of the charges across the homojunction, in which the holes from the *p*-channel were compensated by the electrons brought in by the top gate (similar shifting behavior of $I_{\text{ds}} - V_{\text{tg}}$ curves was also found when the bottom gate voltage increased, as shown in **Figure S4**). The nearly invariable on-off ratio and mobility were consistent with previous reports of dual-gate transistors.[32] The change in threshold voltages, as well as the current, was caused by electrostatic interaction between the bottom and top gate.[40] It was found that the current was dramatically suppressed under the opposite dual-gating, especially in the *p*-branch. To explain this phenomenon, a simplified model was proposed to investigate the electrostatic environment inside the WSe$_2$ channel,[41-43] as shown in **Figure S5** and the relevant statements in Supporting Information. The potential across the WSe$_2$ channel in the vertical direction was numerically obtained by solving the Poisson equation with the continuous conditions of the constant electric displacement at the interfaces.

**Figure 2**c depicts the potential profile along the vertical direction for different top gate voltages $V_{tg}$ at a fixed bottom gating ($V_{bg} = -75$ V). With the help of the gating effect, electrons or holes accumulated at the WSe$_2$ interfaces depending on the polarity of the gating, and the potential therein bended down or bended up accordingly, as shown in **Figure S5**d **and S5**e. For the opposite dual-gating, a vertical *p-n* homojunction was built inside the multilayer WSe$_2$ with an electric field up to ~10$^7$ V/m. The configuration of the dual-channel was also formed along the WSe$_2$ channel: the *n*- and *p*-channel were close to the top and bottom WSe$_2$ interface, respectively. Therefore, conductivity could be evaluated using $\sigma = qn\mu_e + qp\mu_h$ with the assumption of a negligible diffusion current, where $q$ is the elementary charge, and $n$ and $p$ are the free electron and hole concentration respectively. $n$ (in the *n*-channel) and $p$ (in the *p*-channel) increased and decreased when $V_{tg}$ increased, respectively, as shown in **Figure S5**e. However, the mobility measured above was almost invariable with top gating, which implied that the mobility $\mu_h$ of the dominating holes was also invariable. As a result, the current in the *p*-branch was suppressed under the positive top gating due to the decreased hole density. In addition, the built-in electric field inside the *p-n* homojunction was weakened with the increased thickness of WSe$_2$ (**Figure S5**f), mainly due to the screen effect from charge accumulation at the interface.[44]

The built-in field in the *p-n* homojunction helped to separate photogenerated electron-hole pairs, and thus was beneficial for photodetection. We used a 532 nm laser spot, 2 mm diameter, to illuminate the device globally, so as to eliminate the photoresponse contribution from the thermoelectric effect.[45, 46] **Figure 2**d shows the transfer curves at top gate voltages of $V_{tg} = 0$ V and 2 V, where a noteworthy reduction in the dark current was observed for the opposite dual-gating. The photocurrent, defined as the difference between dark current and current under illumination ($I_{ph} = I_{illumination} - I_{dark}$), increased with laser power density ($P_{in}$) from 0 to 1.1 mW/cm$^2$. The photogating effect was confirmed by the shift of the transfer curves, probably arising from charge trapping in the bandgap by traps resulting from disorder, defects, or Se vacancies.[20] Another device with hole mobility $\mu_h$~88 cm$^2$/(V s) at $V_{ds} = 0.1$ V was used for interpreting the detailed photoresponse, characterized briefly in **Figure S6**. The photocurrent increased from zero as the drain voltage increased (as shown in **Figure S6**e and **S6**f). No photocurrent at $V_{ds} = 0$ V implies the absent contribution to the photoresponse from photovoltaic effect. **Figure 2**e depicts the spatially resolved scanning photocurrent image of a device (channel area ~16×8.5 μm$^2$) carried out with a small laser spot (~2 μm-diameter) when $V_{ds} = 0.1$ V and $V_{bg} = -70$ V. It clearly verified that the photocurrent came primarily from the WSe$_2$ channel, rather than from the Schottky junctions between the WSe$_2$ channel and Au electrodes. The parasitic photocurrent outside the WSe$_2$ channel was probably caused by the large size of the laser spot. It is also found $I_{ph}$ at opposite dual-gating ($V_{tg} = 3$ V) is noteworthy larger than that at $V_{tg} = 0$ V. Without drain voltage (i.e., $V_{ds} = 0$ V), no observable photocurrent was found under the opposite dual-gating, as shown in **Figure 2**f, confirming again the absence of photovoltaic current in the device.

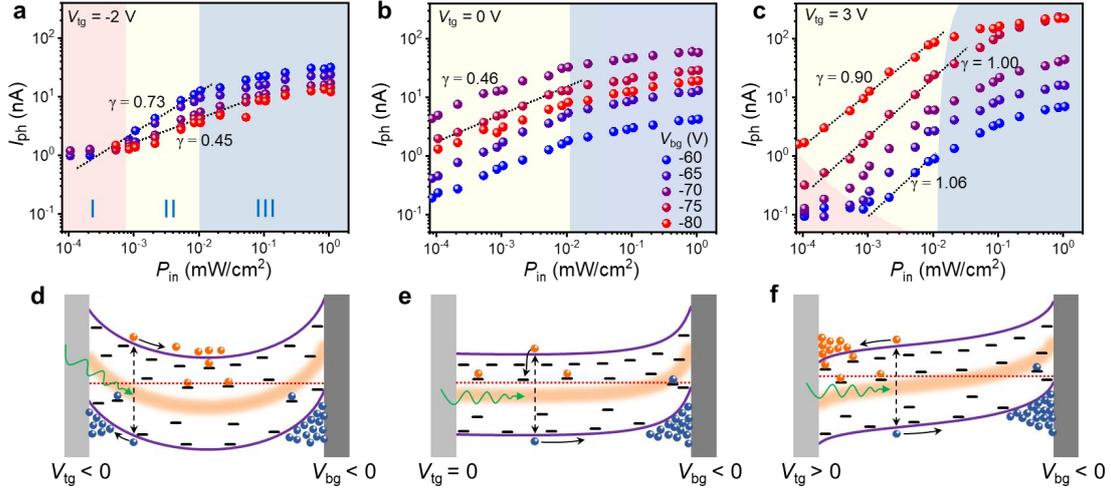

**Figure 3.** Photoresponse characterization of the dual-gate WSe$_2$ photodetector when $V_{ds} = 0.1$ V. **a-c** Logarithmic plots showing the dependence of photocurrent ($I_{ph}$) on laser power density ($P_{in}$) for the top gate voltage $V_{tg} = -2, 0, 3$ V, respectively. Different regimes could be distinguished with colors. **d-f** Schematic illustration of the carrier behavior under illumination in the cross section of the WSe$_2$ channel, where the red/blue sphere indicates electron/hole, the discrete line segments and red dotted lines inside the bandgap depict the trap states and Fermi level, respectively, and the orange regions close to the midgap denote the recombination centers. It should be noted that the Fermi level is also bended with the increased laser intensity.

To get deeper insights into carrier dynamics under illumination, a transient photocurrent method was adopted to record the photoresponse at different gate voltages. When the device was activated by a switchable laser with a suitable modulation frequency (3 Hz in our measurements, chosen based on the photoresponse time of the dual-gate phototransistor), a corresponding square wave of photocurrent was generated. At the given $V_{ds}$ and $V_{tg}$, $V_{bg}$ varied from −80 V to 0 V with a step of 5 V during measurement, where each voltage step was held for several periods of the laser modulation. Finally, the photocurrent at each $V_{bg}$ step was extracted for different laser power densities. The results at $V_{ds} = 0.1$ V are shown in **Figure 3**a–3c and **S7**. For small $V_{ds}$, two or three regimes of photocurrent with different linearities (two for $V_{tg} = 0$ V and three for $V_{tg} \neq 0$ V) were found as the laser intensity increased from $10^{-5}$ to 1 mW/cm². The photocurrent with respect to the laser intensity at each regime could be fitted by $I_{ph} \propto P_{in}^{\gamma}$, and the linear photoresponse occurred if $\gamma \approx 1$. For the regime II when $V_{tg} \neq 0$ V, the linearity could be tuned readily by the top and bottom gate, and the linear case with $\gamma \approx 1$ was attained at the opposite dual-gating. It has been reported that the linearity can be adopted to determine the dominant photoresponse operating mechanism between photoconductive and photogating effects.[46] For our devices, these two effects also coexisted, and were dominant in turn when $\gamma \approx 1$ and $\gamma < 1$, respectively. However, $\gamma > 1$ was found in our experiments (**Figure S7e, S7f**), indicating more complex competitive process between two effects under the illumination that was deeply investigated later.

The input-output efficiency for a photodetector is the responsivity ($R$, in units of A W$^{-1}$), which is defined as the ratio of photocurrent and laser intensity illuminating the active area:[47]

$$R = \frac{I_{ph}}{P_{in}A} \tag{1}$$

where $A$ is the phototransistor channel area. Based on the photocurrent shown in **Figure S7**, the responsivities of the phototransistor were calculated and shown in **Figure S8**. It can be seen that $R$ decreased with laser intensity in both regimes I and III. However, it was possible to tune $R$ in regime II from decreasing to increasing using the dual-gating, and even to make it constant at the opposite dual-gating, namely the linear case with $\gamma \approx 1$. For example, when $V_{tg} = 2$ V and $V_{bg} = -80$ V (**Figure S7e and S8e**), the linear photoresponse was estimated in the intensity regime from $2\times10^{-4}$ to $1\times10^{-2}$ mW/cm², with responsivity up to ~$2.5\times10^4$ A W$^{-1}$. The linear dynamic range, $LDR = 20\log(P_h/P_l)$ in dB for constant responsivity, was about 34 dB, where $P_h$ and $P_l$ are the highest and lowest laser powers respectively of the linear regime. LDR could also be expanded by adjusting the dual-gating: for example it reached 40 dB when $V_{tg} = 3$ V and $V_{bg} = -75$ V. The EQE of a photodetector, expressed as $EQE = R \cdot (hc/q\lambda)$, is generally defined as the number ratio of electron-hole pairs (that contribute to the photocurrent) to the incident photons.[6] The constant $hc/q\lambda$ in our measurements was about 2.33 W A$^{-1}$, and therefore, the EQE at the linear regime was ~5.8×10$^6$ %.

Different photoresponse regimes are probably results of the coaction of a complex process involving carrier generation, separation, recombination, trapping, and conduction, within the same WSe$_2$ channel.[20, 48, 49, 29, 23] The internal photoconductive gain is usually used to reveal the dominant process under illumination, which can be calculated by[50]

$$G = \frac{I_{ph}/q\lambda}{P_{abs}A/hc} = R\frac{hc}{q\lambda\eta_{abs}} \tag{2}$$

when the charge transfer efficiency is assumed to be 100%. $P_{abs} = P_{in}\eta_{abs}$ and $\eta_{abs}$ is the light absorbance.[51, 48] The absorbance can be estimated with the Lambert-Beer law $\eta_{abs} = 1 - e^{\alpha d}$, where $\alpha$ and $d$ are the absorption coefficient and thickness, respectively. It was reported previously that the absorbances of monolayer WSe$_2$ and graphene for 532 nm are about 4% and 2.3%[52, 53] respectively, leading to the actual absorbance of $\eta_{abs}$~43.5% by the WSe$_2$ channel. Therefore, the photoconductive gain $G$~5.47$R$ was found in our measurements and $G$ at the linear regime reached up to ~1.34×10$^5$ when $V_{tg} = 2$ V and $V_{bg} = -80$ V. The photoconductive gain can also be estimated by the carrier lifetime ($\tau_l$) and the majority transit time $\tau_t = L^2/\mu V_{ds}$, i.e., $G = \tau_l/\tau_t$. The carrier lifetime is directly related to the trap lifetime. At low intensity (regime I), the longer-lived and deeper-lying traps are gradually filled by photogenerated carriers,[49] where the gain becomes very large and gradually reduces with the intensity owing to the shorter lifetime of the shallow traps to be filled. Once the traps are almost filled (regime II), the separation of photogenerated electron-hole pairs by the strong built-in field dominates. The separated electrons or holes accumulate into the relevant channels gradually with increasing laser intensity, resulting in a growing screen effect against the gate voltages. The bending of the potential profile is offset and weakens the built-in field, which prolongs the lifetime of the traps[54] and improves the photoconductive gain as well. The competitive process of gain decrease (from shallow traps filling) and gain increase (from charge accumulation) produces the tunable linearity of the photoresponse. When the laser intensity increases further (regime III), the channels tend to be saturated with continuing charge accumulation, and the photoconductive gain decreases significantly. For the specific case at $V_{tg} = 0$ V, regime II disappears because of the absence of the *p-n* homojunction inside the WSe$_2$ channel.

Based on the calculated potential profile (**Figure 2**c), the schematic band structures of WSe$_2$ in the vertical direction at $V_{bg} < 0$ V were also adopted to illustrate these regimes at different values of $V_{tg}$, as shown in **Figure 3**d–3f. When $V_{tg} < 0$ V, the band structure bended upwards at both the top and bottom interfaces of the WSe$_2$ channel and formed two *p*-channels there. The resulting built-in field caused the holes to drift towards the *p*-channels and the electrons towards the intermediate region between them. The lifetime of traps was prolonged by the offset of band bending from charge accumulation. The intermediate region was rapidly saturated by the accumulated electrons as the laser intensity increased, exhibiting gain decrease during the competitive process, and finally produced the sublinear photoresponse ($\gamma < 1$, see regime II in **Figure 3**a). When $V_{tg} = 0$ V, the band bended upwards only at the bottom interface, while the band away from the bottom interface was flat. The abundant deep traps in the flat band led to a very large gain under weak illumination. However, the offset of band bending due to charge accumulation hardly arose with the absence of the *p-n* homojunction, causing regime II in **Figure 3**b to vanish. If $V_{tg} > 0$ V, the band structure bended upwards at the bottom interface and downwards at the top interface, forming a typical *p-n* homojunction in the vertical direction. The Fermi level in the *p/n*-channel was close to the valence/conductance band, indicating that some traps were already filled by the holes/electrons there. Thus, the photoconductance gain at low laser intensity (regime I) was much smaller than when $V_{tg} < 0$ V and $V_{tg} = 0$ V, as shown in **Figure 4**a. With the increase of laser intensity, the charge accumulation gradually dominated, resulting in the offset of band bending as well as the increased photoconductive gain. Regime II in **Figure S7**e, **S7**f, and **S9** shows clearly the transition from $\gamma > 1$ to $\gamma < 1$ when $V_{bg}$ increased from −60 V to −80 V. The charge accumulation in the *p/n*-channel was able to boost the photocurrent through the photoconductive effect, which was faster than the photogating effect. As a result, the response became faster as the laser intensity increased. **Figure S10** shows the rise time (~60 ms) and decay time (~100 ms), measured between 10% and 90% of the maximum photocurrent, in regime II when $V_{bg} = -75$ V and $V_{tg} = 3$ V, which both decreased to < 5 ms in regime III.

Another important measure of photodetector sensitivity is noise equivalent power (NEP, units W Hz$^{-1/2}$). It is defined as the input signal power at which the signal-to-noise ratio is 1 and the output bandwidth is 1 Hz.[55] A lower NEP means the detector is more sensitive, and the minimum NEP is often quoted as a measure of the sensitivity of a photodetector. For our dual-gate WSe$_2$ phototransistor, NEP can be expressed as $NEP = \left(\sqrt{\langle i_n^2 \rangle / \Delta f}\right)/R$, where $\Delta f$ is the measurement bandwidth over which the noise is considered. Without any output filtering, the photodetector's bandwidth is well approximated by the measurement bandwidth. The noise current is calculated using $\langle i_n^2 \rangle = \int_0^{\Delta f} S(f) df$, where $S(f)$ is the spectral noise density (measured under dark conditions) with respect to the noise frequency $f$, as shown in **Figure 4**b and **S11**. Note that $NEP \sim \sqrt{S(f)}/R$ for a small measurement bandwidth. Two main sources of noise are observed, including $1/f$ noise and $f$-independent white noise. $1/f$ noise (so-called flicker noise, which decreases with increasing frequency), is always observed in transistors because of carrier capture and release.[56-58] The white noise in electronic circuits usually includes temperature-dependent Johnson-Nyquist noise (or thermal noise) and discrete charge induced shot noise. These two sources of white noise cannot easily be distinguished by experimental observation. However, shot noise may become dominant at high frequencies and low temperatures if the electric current is very small, or even be negligible when $V_{ds} = 0$ V.[59] Our results found that $1/f$ noise increased under homogeneous dual-gating because of the additional noise source from the new channel. However, for the opposite dual-gating, carrier capture and release are lessened owing to the filling of traps in the *n*-channel and *p*-channel, as shown in Figure 3f, leading to very low $1/f$ noise. The low $1/f$ noise is highly advantageous for NEP. For example, when $V_{tg} = 3$ V and $V_{bg} = -75$ V, NEP in the linear photoresponse regime is estimated at 3 fW Hz$^{-1/2}$, which is one order of magnitude smaller than that of commercial silicon avalanche photodiodes (30 fW Hz$^{-1/2}$).[60]

Notice that NEP is not concerned with the active area of a photodetector. The detectivity $D^*$, in units of cm Hz$^{1/2}$ W$^{-1}$ or Jones, is defined as the inverse of NEP normalized to the square root of active area ($A$). It is introduced to evaluate the comparable signal-to-noise ratio of a photodetector with variable active area:

$$D^* = \frac{\sqrt{A}}{NEP} = \frac{R\sqrt{A\Delta f}}{\sqrt{\int_0^{\Delta f} S(f) df}} \quad (3)$$

For a small bandwidth,[23] the detectivity can be simplified as $D^* = R\sqrt{A/S(f)}$. **Figure 4**c depicts the obtained detectivity of our device at different values of $V_{tg}$ when $V_{bg} = -75$ V. It is shown that the detectivity was greatly enhanced for the opposite dual-gating, and reached a plateau value up to $2\times10^{13}$ Jones in the linear regime, even though the responsivity was not at the optimum. It should be noted that the detectivity of our phototransistor was higher than that of commercial silicon photodetectors (500–600 mA W$^{-1}$, ~$10^{12}$ Jones, 200–1100 nm).[61, 23] The main reason for this high detectivity is the noteworthy reduction of the $1/f$ noise under opposite dual-gating.

After being kept in an ordinary plastic vacuum tank for nine months, the dual-gate WSe$_2$ phototransistor was measured again in regime III with a switchable laser (modulation frequencies 200, 500, 1000, 2000 Hz), as shown in **Figure S12** and **4**d. The results showed that the top gate voltage was still beneficial for the photocurrent and the signal-to-noise ratio.

The good reproducibility and stability implied excellent encapsulation properties of hBN. When $V_{tg} = 3$ V and $V_{bg} = -75$ V, the rise time (~0.47 ms) and the decay time (~1.56 ms), measured between 10% and 90% of the maximum photocurrent, were consistent with the response time of the fast decay process of the original device shown in **Figure S10**b and **S10**d. At laser modulation frequency 2000 Hz, the photocurrent signal was still distinguishable under the opposite dual-gating (**Figure S12**d), implying that it retained the capability of high-frequency photodetection. Compared to phototransistors with a linear photoresponse reported elsewhere, our dual-gate phototransistors exhibited many superiorities attributed to the layered 2D material and device configuration. Our devices exhibited detectivity three orders of magnitude higher than a HfO$_2$-encapsulated MoS$_2$ phototransistor (~2×10$^{10}$ Jones),[20] the speed three orders faster than a dual-gate MoS$_2$ phototransistor (with photoresponse time > 50 s),[22] and to be more stable than perovskite-based[21] and organic-based[23] phototransistors. For the dual-gate MoS$_2$ phototransistor with similar architecture,[22] only *n*-channel can be turned on and the photoresponse was boosted by enhancing the photogating effect through the interface coupling, which would be weakened for thick MoS$_2$ film. Moreover, the linear photoresponse appeared when the *n*-channel was turned off. However, *p*-channel and *n*-channel of our devices can be turned on simultaneously because of the thick and ambipolar WSe$_2$ film. The linear photoresponse with high detectivity was realized at the "on" state (i.e., channels were turned on) by the complex competitive process between photoconductive effect and photogating effect, benefiting from the built-in vertical *p-n* homojunction.

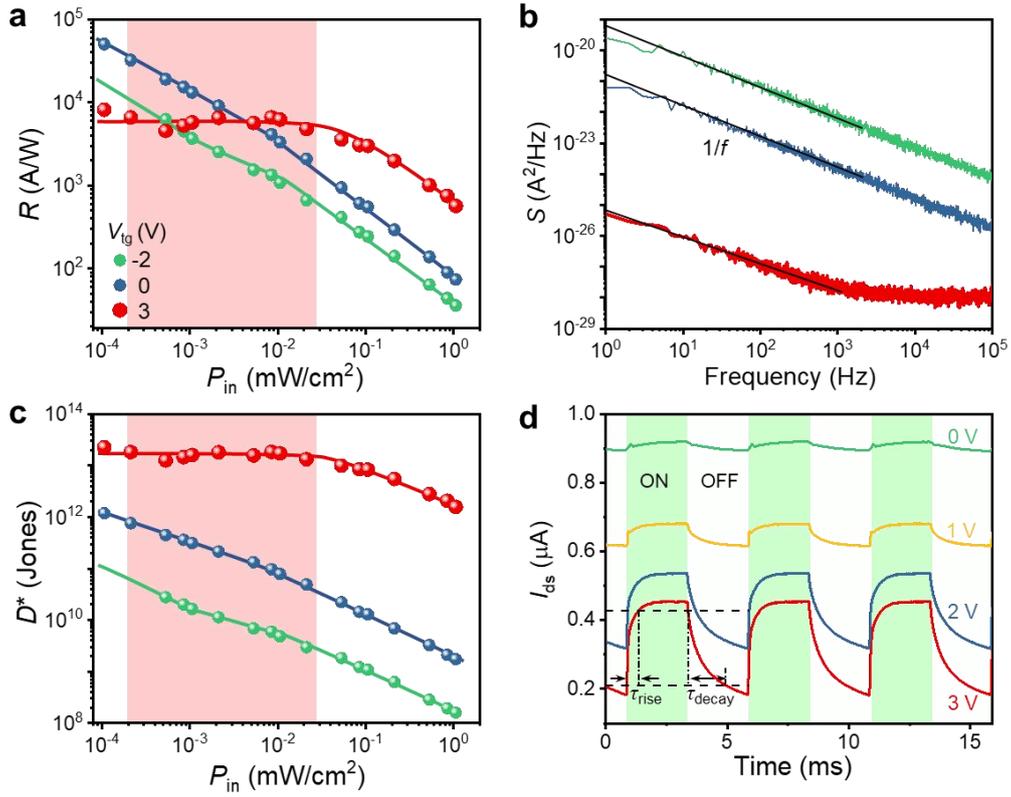

**Figure 4.** Optoelectronic properties of the dual-gate WSe$_2$ photodetector at the fixed $V_{ds} = 0.1$ V and $V_{bg} = -75$ V. **a** Logarithmic plot of the responsivity with respect to the laser intensity at different $V_{tg}$. **b** and **c** are the corresponding noise current spectral density $S(f)$ and detectivity $D^*$. The light red colored regimes in **a** and **c** denote the linear photoresponse regimes at $V_{tg} = 2$ V. **d** The transient photoresponse of the device after nine months for the laser power density ~ 1.0 mW cm$^{-2}$. The modulation frequency of the switchable laser is set as 200 Hz.

## 3. Conclusions

We fabricated dual-gate WSe$_2$ phototransistors with the dry transfer technique, where a multilayer ambipolar WSe$_2$ channel was sandwiched with hBN nanosheets and a uniform top gate was made from transparent graphene. The vertical electric field inside the WSe$_2$ channel was tunable using the dual-gating, resulting in charge accumulation at both WSe$_2$ interfaces. Dual-channel conduction was formed under the opposite dual-gating, where the built-in vertical *p-n* homojunction was highly effective at separating photogenerated electron-hole pairs and also reduced the 1/f noise of the device. The photoconductive gain was tuned to be constant by adjusting the top and bottom gating, which is the key for a linear photoresponse. From our results, a linear photoresponse with high responsivity of ~$2.5\times10^4$ A W$^{-1}$ and high detectivity of ~$2\times10^{13}$ Jones was realized. Compared with state-of-the-art WSe$_2$ based photodetectors, as shown in **Table S1**, our vertical dual-gate WSe$_2$ phototransistors exhibited excellent detection performance with a tunable linear photoresponse regime. The linear photoresponse has the potential to be further optimized by adjusting the dual-gating, and the large gate voltage could be reduced by increasing the dielectric capacitance, for example by thinning the high-quality dielectric layer or using higher-κ dielectrics. In short, our dual-gate WSe$_2$ phototransistors with tunable linear photoresponse and high-performance provide a promising strategy for high-resolution and quantitative light detection with layered 2D semiconductors.

## 4. Experimental Section

*Fabrication*: All the multilayer WSe$_2$, hBN, and graphene nanosheets were exfoliated from commercially available bulk crystals (HQ Graphene Company) by micromechanical cleavage technique with Scotch-tape. The dual-gate phototransistor was then fabricated by the dry transfer technique with the assistance of polydimethylsiloxane (PDMS) stamp. First, hBN nanosheets were deposited onto the pre-cleaned Si/SiO$_2$ substrate (highly *p*-doped Si with resistivity < $5\times10^{-3}$ Ω cm, the SiO$_2$ thickness ~ 285 nm). WSe$_2$ nanosheets and Au (thickness

~ 50 nm) electrodes were then transferred in sequence on the hBN to form the channel and source/drain electrode. After that, another hBN nanosheet was transferred to cover on the whole WSe$_2$ channel as the dielectric and encapsulation layer. Following that, the whole WSe$_2$ channel was covered with a large area graphene (i.e., the top gate). Finally, an Au (thickness ~ 50 nm) electrode was transferred onto the graphene as the top gate electrode. The as-prepared device was annealed at 180 °C in a vacuum tube furnace for 0.5 hours to remove the resisted residues and to decrease contact resistance. The fabrication process is also shown in Figure S1.

*Characterization*: Raman and PL spectra of WSe$_2$ flakes were collected by using a confocal mirco-Raman system (Alpha300R, WITec) excited by 532 nm laser (spot size ~ 400 nm, laser power = 1.0 mW, resolution ~ 0.02 cm$^{-1}$). AFM images were obtained by an atomic force microscope (Dimension Icon, Bruker). All the electrical/optoelectrical measurements were carried out at room temperature by a semiconductor parameter analyzer (PDA FS380 Pro, Platform Design Automation) in a probe station with the vacuum degree of 6×10$^{-6}$ mbar. The photocurrent is excited with a 532 nm laser (diameter ~ 2 mm).

**Supporting Information**

Supporting Information is available from the Wiley Online Library or from the author.

**Acknowledgements**

Jinpeng Xu and Xiaoguang Luo contributed equally to this work. This work was supported by the National Natural Science Foundation of China (NSFC) (Grant Nos. 61905198, 61575094), the Natural Science Basic Research Program of Shaanxi Province (Program No. 2019JQ-059), the National Postdoctoral Program for Innovative Talents (No. BX20190283), the Joint Research Funds of Department of Science & Technology of Shaanxi Province and Northwestern Polytechnical University (Nos. 2020GXLH-Z-020, 2020GXLH-Z-027 and 2020GXLH-Z-029), and the Fundamental Research Funds for the Central Universities. We also thank the Analytical & Testing Center of NPU for the assistance in device fabrication.

Received: ()
Revised: ()
Published online: ()